\documentclass[%
    reprint,
  superscriptaddress,
  amsmath,
  amssymb,
  aps,
  pra,
]{revtex4-2}

\usepackage{dcolumn}
\usepackage{bm}
\usepackage{braket}
\usepackage{graphicx}
\usepackage{float}
\usepackage{tikz}
\usepackage{comment}
\usetikzlibrary{arrows.meta,decorations.pathreplacing,calc}
\usepackage[hidelinks]{hyperref}

\begin{document}

\preprint{APS/123-QED}

\title{SWAP Gate Between Polarization and Time-Bin Qubits on a Single Photon}

\author{Adam Sultan}
\email{adamsultan@cmail.carleton.ca}
\author{Connor Kupchak}
\email{ConnorKupchak@cunet.carleton.ca}
\affiliation{Carleton University, Ottawa, Ontario, Canada}

\begin{abstract}

Multi-degree-of-freedom photonic quantum processing requires coherent routing between qubit encodings on a single photon. We propose a SWAP circuit between polarization and time-bin qubits using polarization-dependent delays and synchronized electro-optic modulation, analyzing phase-calibration, voltage-accuracy, and timing requirements for experimental realization. The architecture follows a three-CNOT SWAP decomposition, using two polarization beam splitter (PBS)--delay--PBS stages to produce polarization-controlled time-bin parity flips. Between these stages, an electro-optic modulator (EOM) driven by a voltage waveform synchronized to a time-bin grid flips polarization in odd bins while leaving even bins unchanged. A final PBS--delay--PBS compensation stage delays the H-polarized components by $2\Delta t$, aligning the relevant H and V amplitudes in common output time bins. We describe an infinite grid of repeating early/late time-bin pairs, labelling early bins as even ($\ket{E}$) and late bins as odd ($\ket{O}$). Throughout, we use these even/odd labels without explicit pair indices. A delay of one bin spacing $\Delta t$ moves an even bin to the following odd bin and an odd bin to the following even bin, corresponding to the early slot of the next pair in a traditional early/late encoding. Tracking the polarization and time-bin states through the complete circuit, we establish the resulting SWAP transformation. We identify the optical phase conditions required for coherent operation and discuss their calibration and compensation. We also quantify the effect of EOM voltage errors on polarization-flip probability and present an illustrative timing budget based on switching, detector resolution, and timing jitter.

\end{abstract}
\maketitle

\section{Introduction}

Linear-optical architectures are among the simplest photonic platforms to implement, but their scalability remains challenging \cite{ref_humphreys_linear}. One approach to addressing this challenge is to encode additional information in multiple degrees of freedom on a single photon, accessing higher-dimensional Hilbert spaces, while standard optical components can be used to manipulate these states~\cite{ref_dhand_unitary,ref_su_hybrid,ref_klm}. Such architectures can encode qubits in polarization, arrival time, and spatial mode, enabling compact implementations of multi-qubit operations \cite{ref_multiDOF_toffoli}. Time-bin and polarization encodings are particularly well suited to optical communication and transmission through different media \cite{ref_bussieres_unitary}.

In this work, the two qubit ``wires'' represent two two-state subsystems of the same photon rather than two separate photons. Polarization provides one qubit, $(\ket{H},\ket{V})$, while time-bin parity provides the other, $(\ket{E},\ket{O})$. The SWAP operation coherently exchanges the quantum states of these two subsystems, transferring information encoded in polarization to the time-bin degree of freedom and vice versa. This inter-degree-of-freedom operation provides a routing primitive for connecting different encodings within a photonic circuit.

Inter-degree-of-freedom SWAP operations have been experimentally demonstrated by implementing a single-photon SWAP gate between momentum and polarization, using the operation to swap information between two distinct degrees of freedom \cite{ref_fiorentino_swap}. Their work demonstrates the utility of a SWAP gate as a coherent routing primitive within a single photon, although the exchanged degrees of freedom were momentum and polarization rather than time and polarization.
Transfer between time-bin and polarization encodings has also been demonstrated using an ultrafast optical Kerr shutter \cite{ref_kupchak_conversion}.

A polarization--time-bin SWAP gate was also proposed by Zhang \emph{et al.} as a component of a deterministic nondestructive hyperentangled Bell-state analyzer~\cite{ref_zhang_hyperentangled_analysis}. Their three-CNOT architecture uses PBS-based conditional routing, time-bin-flip operations, and a Pockels cell for the time-bin-controlled polarization flip. Building on the polarization--time-bin SWAP proposed by Zhang \emph{et al.}~\cite{ref_zhang_hyperentangled_analysis}, we present a new circuit in which fixed polarization-dependent delays produce bidirectional changes in logical time-bin parity, while physical propagation remains forward in time. Synchronized electro-optic modulation and a final temporal-compensation stage combine these operations into a coherent polarization--time-bin SWAP under the stated ideal operating and phase conditions. We outline the phase-calibration and timing requirements relevant to experimental realization.

The remainder of this paper is organized as follows. Section~\ref{sec:encoding} defines the logical qubit encodings and the parity-based time-bin grid. Section~\ref{sec:swap} presents the three-CNOT SWAP decomposition and a basis-state proof. Section~\ref{sec:implementation} maps each CNOT to standard optical components. Section~\ref{sec:constraints} discusses EOM polarization-rotation error probability and timing-resolution requirements for experimental realization.

\section{System Definition and Encoding}
\label{sec:encoding}

We define the polarization and time-bin qubit encodings and the parity-based time-bin grid that enables the physical CNOT operations developed below.

\subsection{Polarization and Time-Bin Qubits}

The polarization bases are defined as Horizontal and Vertical using the standard notation: $\ket{H}, \quad \ket{V}$ and the time-bin basis as even and odd multiples of a time-bin spacing $\Delta t$: $\ket{E}, \quad \ket{O}$. For compact notation, we identify the logical basis states as $\ket{0}_P\equiv\ket{H}$, $\ket{1}_P\equiv\ket{V}$, $\ket{0}_T\equiv\ket{E}$, and $\ket{1}_T\equiv\ket{O}$. The total Hilbert space is $\mathcal{H}=\mathcal{H}_{pol}\otimes\mathcal{H}_{time}$.

A single-photon state in the two-qubit Hilbert space can therefore be written as

\begin{equation}
\ket{\psi} = (\alpha\ket{H} + \beta\ket{V}) \otimes (\gamma\ket{E} + \delta\ket{O}),
\end{equation}

Here, $\alpha$ and $\beta$ are the probability amplitudes for the polarization states $\ket{H}$ and $\ket{V}$, respectively, while $\gamma$ and $\delta$ are the probability amplitudes for the time-bin states $\ket{E}$ and $\ket{O}$ with $|\alpha|^2 + |\beta|^2 = 1$ and $|\gamma|^2 + |\delta|^2 = 1$.

\subsection{Even and Odd Time-Bin Labels}

We describe an infinite grid of repeating early/late time-bin
pairs separated by $\Delta t$. Early bins are labelled even
($\ket{E}$), and late bins are labelled odd ($\ket{O}$).
Throughout, we use these even/odd labels without explicit
pair indices.

A delay of $\Delta t$ moves an even component into the
following odd bin and an odd component into the following
even bin. In early/late terminology, the latter corresponds
to moving a late component into the early slot of the next
pair. The delay therefore changes the parity label while
always advancing the physical arrival time.

Bins with the same parity remain distinct temporal modes.
The even/odd notation does not imply that amplitudes in
different physical bins overlap. We therefore track their
arrival times through the circuit and use a final
polarization-dependent delay to bring the relevant
polarization amplitudes into the same output bins.

The periodic grid also provides a timing reference for the
EOM voltage waveform. The drive is synchronized so that
odd bins experience a polarization flip while even bins
remain unchanged. The photon wavepackets must lie within
the stable portions of the modulation waveform.

Figure~\ref{fig:parity-grid} illustrates the alternating
even and odd bins and the action of a single-bin delay.

\begin{figure}[H]

\centering

\resizebox{0.85\columnwidth}{!}{%
\begin{tikzpicture}[x=1.35cm,y=1cm,>=Latex]

    \draw[->] (-0.1,0) -- (8.6,0) node[right] {$t$};

    \fill[blue!10] (0,0) rectangle (2,1.15);
    \fill[orange!12] (2,0) rectangle (4,1.15);
    \fill[blue!10] (4,0) rectangle (6,1.15);
    \fill[orange!12] (6,0) rectangle (8,1.15);

    \foreach \x in {0,2,4,6,8} {
        \draw[densely dashed,gray] (\x,0) -- (\x,1.15);
    }

    \node at (1,0.58) {\small even ($\ket{0}_T$)};
    \node at (3,0.58) {\small odd ($\ket{1}_T$)};
    \node at (5,0.58) {\small even ($\ket{0}_T$)};
    \node at (7,0.58) {\small odd ($\ket{1}_T$)};

    \node[below] at (0,0) {\small $t_0$};
    \node[below] at (2,0) {\small $t_0+\Delta t$};
    \node[below] at (4,0) {\small $t_0+2\Delta t$};
    \node[below] at (6,0) {\small $t_0+3\Delta t$};
    \node[below] at (8,0) {\small $t_0+4\Delta t$};

    \draw[decorate,decoration={brace,amplitude=5pt}] (0,1.3) -- (2,1.3)
        node[midway,yshift=9pt] {$\Delta t$};

    \draw[densely dotted,gray] (0.95,1.15) -- (0.95,2.05);
    \draw[densely dotted,gray] (3.15,1.15) -- (3.15,2.05);

    \fill[black] (0.95,2.25) circle (1.5pt);
    \node[above] at (0.95,2.4) {\small photon};

    \draw[->,thick,red] (1.2,2.0) -- (2.9,2.0)
        node[midway,below,yshift=-2pt] {\small $+\Delta t$ delay};

    \fill[black] (3.15,2.25) circle (1.5pt);
    \node[above] at (3.15,2.4) {\small photon};

    \draw[->,thin] (1.1,2.3) to[out=15,in=165] (3.0,2.3);

\end{tikzpicture}%
}

\caption{Parity-based time-bin grid divided into width-$\Delta t$ segments.}
\label{fig:parity-grid}

\end{figure}

Figure~\ref{fig:eom-sampling} shows the EOM voltage waveform synchronized with photon arrival times, where even and odd time bins correspond to $\ket{0}_T$ and $\ket{1}_T$, respectively.

\begin{figure}[H]

\centering

\resizebox{\columnwidth}{!}{%

\begin{tikzpicture}[x=1.45cm,y=1.35cm,>=Latex]

    \draw[->] (0.3,0.0) -- (9.6,0.0) node[right] {$t$};

    \draw[->] (0.3,2.2) -- (0.3,4.0) node[above] {EOM voltage};

    \draw[very thick] (0.7,2.45) -- (2.0,2.45) -- (2.0,3.75) -- (3.3,3.75) -- (3.3,2.45) -- (4.6,2.45) -- (4.6,3.75) -- (5.9,3.75) -- (5.9,2.45) -- (7.2,2.45) -- (7.2,3.75) -- (8.5,3.75) -- (8.5,2.45) -- (9.0,2.45);

    \foreach \x in {0.7,2.0,3.3,4.6,5.9,7.2,8.5} {

        \draw[densely dashed] (\x,2.25) -- (\x,3.85);

        \draw[fill=black] (\x,1.0) circle (1.3pt);

        \draw[->,thin] (\x,1.0) -- (\x,2.6);

    }

    \node[below] at (0.7,0.95) {$t_0$};

    \node[below] at (2.0,0.95) {$t_0+\Delta t$};

    \node[below] at (3.3,0.95) {$t_0+2\Delta t$};

    \node[below] at (4.6,0.95) {$t_0+3\Delta t$};

    \node[below] at (5.9,0.95) {$t_0+4\Delta t$};

    \node[below] at (7.2,0.95) {$t_0+5\Delta t$};

    \node[below] at (8.5,0.95) {$t_0+6\Delta t$};

    \node[below] at (1.35,-0.25) {$T=\ket{E}$};

    \node[below] at (2.65,-0.25) {$T=\ket{O}$};

    \node[below] at (3.95,-0.25) {$T=\ket{E}$};

    \node[below] at (5.25,-0.25) {$T=\ket{O}$};

    \node[below] at (6.55,-0.25) {$T=\ket{E}$};

    \node[below] at (7.85,-0.25) {$T=\ket{O}$};

    \draw[decorate,decoration={brace,amplitude=5pt}] (0.7,4) -- (2.0,4)

        node[midway,yshift=11pt] {$\Delta t$};

    \node[align=center] at (4.8,4.15) {stable sample points};

    \draw[->,thin] (3.5,3.7) -- (2.65,3.45);

\end{tikzpicture}%

}

\caption{EOM timing diagram for the parity-based time-bin encoding.}
\label{fig:eom-sampling}

\end{figure}

\section{SWAP Gate Composition}
\label{sec:swap}

The SWAP gate can be decomposed into three sequential CNOT operations, a standard circuit identity in quantum computing \cite{ref_nielsen_chuang,ref_barenco_cnot}:

\begin{equation}
\text{SWAP} = \mathrm{CNOT}_{P \rightarrow T}\,\mathrm{CNOT}_{T \rightarrow P}\,\mathrm{CNOT}_{P \rightarrow T}
\label{eq:swap-ptp}
\end{equation}

\begin{figure}[H]

\centering

\resizebox{0.98\columnwidth}{!}{%
\begin{tikzpicture}[x=1cm,y=1cm,>=Latex]

    \node[left] at (0,1.2) {$P$};
    \node[left] at (0,0) {$T$};

    \draw[thick] (0,1.2) -- (8.2,1.2);
    \draw[thick] (0,0) -- (8.2,0);

    \fill (1.8,1.2) circle (2.2pt);
    \draw (1.8,0) circle (0.34);
    \draw (1.8,-0.34) -- (1.8,0.34);
    \draw (1.46,0) -- (2.14,0);
    \draw[thick] (1.8,1.2) -- (1.8,0);

    \fill (4.1,0) circle (2.2pt);
    \draw (4.1,1.2) circle (0.34);
    \draw (4.1,0.86) -- (4.1,1.54);
    \draw (3.76,1.2) -- (4.44,1.2);
    \draw[thick] (4.1,0) -- (4.1,1.2);

    \fill (6.4,1.2) circle (2.2pt);
    \draw (6.4,0) circle (0.34);
    \draw (6.4,-0.34) -- (6.4,0.34);
    \draw (6.06,0) -- (6.74,0);
    \draw[thick] (6.4,1.2) -- (6.4,0);

    \node at (9.2,0.6) {$=$};

    \node[left] at (10.2,1.2) {$P$};
    \node[left] at (10.2,0) {$T$};
    \draw[thick] (10.2,1.2) -- (13.8,1.2);
    \draw[thick] (10.2,0) -- (13.8,0);

    \draw[thick] (11.7,1.05) -- (12.1,1.35);
    \draw[thick] (11.7,1.35) -- (12.1,1.05);

    \draw[thick] (11.7,-0.15) -- (12.1,0.15);
    \draw[thick] (11.7,0.15) -- (12.1,-0.15);

    \draw[thick] (11.9,1.2) -- (11.9,0);

\end{tikzpicture}
}%

\caption{Logical SWAP decomposition between polarization ($P$) and time-bin ($T$) qubits: the three-CNOT sequence is equivalent to the canonical crossed-$X$ SWAP symbol ~\cite{ref_nielsen_chuang}.}
\label{fig:swap-circuit}

\end{figure}

Alternatively, the CNOT gates can be rearranged as:

\begin{equation}
\text{SWAP} = \mathrm{CNOT}_{T \rightarrow P}\,\mathrm{CNOT}_{P \rightarrow T}\,\mathrm{CNOT}_{T \rightarrow P}
\end{equation}

Both orderings are feasible implementations of the SWAP operation, however we will focus on the first ordering in Eq.~\eqref{eq:swap-ptp} for the basis-state proof and physical implementation sections. The second ordering can be implemented by rearranging the sequence of optical components accordingly, but it would requre a much more complex time bin compensation stage to ensure that the polarization amplitudes are recombined in the same time bin.

The controlled-NOT operations act on the computational basis as

\begin{align}
\mathrm{CNOT}_{P \rightarrow T}:\quad &\ket{p}_P\ket{t}_T \mapsto \ket{p}_P\ket{t \oplus p}_T, \\
\mathrm{CNOT}_{T \rightarrow P}:\quad &\ket{p}_P\ket{t}_T \mapsto \ket{p \oplus t}_P\ket{t}_T,
\end{align}

\noindent where $\oplus$ denotes the XOR operation.

\subsection{Basis-State Proof of the SWAP}

This proof describes the ideal logical three-CNOT decomposition. In the optical implementation, the physical arrival times must also be tracked, and the final compensation stage is required to obtain the factorized output in common temporal modes.

We will represent $\alpha$ and $\beta$ as the probability amplitudes for the polarization states $\ket{H}$ and $\ket{V}$, respectively, while $\gamma$ and $\delta$ are the probability amplitudes for the time-bin states $\ket{E}$ and $\ket{O}$ with $|\alpha|^2 + |\beta|^2 = 1$ and $|\gamma|^2 + |\delta|^2 = 1$.

We use the logical basis states $\ket{E}$ and $\ket{O}$ for the time-bin qubit: $\ket{0}_P \equiv \ket{H}$, $\ket{1}_P \equiv \ket{V}$, $\ket{0}_T \equiv \ket{E}$, and $\ket{1}_T \equiv \ket{O}$.

Our general single-photon state in the two-qubit Hilbert space is:
\begin{equation}
\ket{\psi_0} = (\alpha\ket{H} + \beta\ket{V}) \otimes (\gamma\ket{E} + \delta\ket{O})
\end{equation}

\noindent\small

\begin{equation}
{\scriptsize
\begin{aligned}
\scriptstyle
\ket{\psi_0} &= \alpha\gamma\ket{HE} + \alpha\delta\ket{HO} + \beta\gamma\ket{VE} + \beta\delta\ket{VO} \\
\mathrm{CNOT}_{P \rightarrow T}:\!\ket{\psi_1} &= \alpha\gamma\ket{HE} + \alpha\delta\ket{HO} + \beta\gamma\ket{VO} + \beta\delta\ket{VE} \\
\mathrm{CNOT}_{T \rightarrow P}:\!\ket{\psi_2} &= \alpha\gamma\ket{HE} + \alpha\delta\ket{VO} + \beta\gamma\ket{HO} + \beta\delta\ket{VE} \\
\mathrm{CNOT}_{P \rightarrow T}:\!\ket{\psi_3} &= \alpha\gamma\ket{HE} + \alpha\delta\ket{VE} + \beta\gamma\ket{HO} + \beta\delta\ket{VO} \\
&= (\gamma\ket{H} + \delta\ket{V}) \otimes (\alpha\ket{E} + \beta\ket{O}).
\end{aligned}
}
\label{eq:swap-state-proof}
\end{equation}

In the computational basis, the sequence implements the standard SWAP operator \cite{ref_nielsen_chuang},
\begin{equation}
U_{\mathrm{SWAP}} =
\begin{pmatrix}
1 & 0 & 0 & 0 \\
0 & 0 & 1 & 0 \\
0 & 1 & 0 & 0 \\
0 & 0 & 0 & 1
\end{pmatrix},
\end{equation}
\noindent mapping $\ket{p}_P\ket{t}_T \mapsto \ket{t}_P\ket{p}_T$.
Since each CNOT factor is unitary, $U_{\mathrm{SWAP}}$ is unitary.

\section{Physical Implementation}
\label{sec:implementation}

\subsection{$\mathrm{CNOT}_{T \rightarrow P}$: Electro-Optic Modulator}

A time-dependent voltage applied to an electro-optic modulator induces a polarization rotation via the Pockels effect. By synchronizing the voltage waveform with photon arrival times, different time bins experience distinct polarization transformations.

An optical Kerr switching device provides an alternative route to the same time-conditioned polarization operation. This can be considered when the insertion loss of an EOM or imperfections in its voltage-controlled modulation would otherwise distort the photonic quantum state. Ultrafast Kerr switching of photonic entanglement has been demonstrated experimentally~\cite{ref_hall_kerr}.

If photon arrivals are indexed by $t_n = t_0 + n\Delta t$, the EOM drive is synchronized to the same clock interval $\Delta t$ so each photon time bin samples the intended voltage phase. Quantitative bounds on $\tau_{switch}$ and $\tau_{open}$ are given in Sec.~\ref{sec:constraints}.

The EOM implements polarization flips when an applied voltage induces a voltage-controlled birefringence so the device behaves as a tunable waveplate. The phase delay between the crystal axes is $\phi = \pi\frac{V}{V_{\pi}}$, so driving the EOM at $V=V_{\pi}$ produces a half-wave delay. With the crystal axes oriented at $45^{\circ}$ to the $H/V$ basis, a half-wave delay reflects the polarization about the crystals horizontal axis, exchanging $H\leftrightarrow V$ (i.e., the polarization vector's H and V amplitudes become swapped).

The EOM implements a time-controlled polarization change in the form:

\begin{equation}
\begin{aligned}
\ket{E}\ket{H} &\mapsto \ket{E}\ket{H}, \\
\ket{O}\ket{H} &\mapsto \ket{O}\ket{V}, \\
\ket{E}\ket{V} &\mapsto \ket{E}\ket{V}, \\
\ket{O}\ket{V} &\mapsto \ket{O}\ket{H},
\end{aligned}
\label{eq:eom-cnot}
\end{equation}

\subsection{$\mathrm{CNOT}_{P \rightarrow T}$: Polarization-Dependent Delay}

A polarization beam splitter (PBS) routes horizontal and vertical components into paths with different lengths, introducing a delay $\Delta t$ for one polarization. Upon recombination, the photon's time bin qubit is conditioned on the photon's polarization. The PBS-delay-PBS system implements a polarization-controlled delay  shown in Fig.~\ref{fig:pbs-delay}.

\begin{figure}[H]

\centering

\resizebox{0.9\columnwidth}{!}{%
\begin{tikzpicture}[>=Stealth, every node/.style={font=\small}, line width=0.9pt]

    \node (in) at (0,0) {Photon in};
    \node[draw,rectangle,minimum width=1.0cm,minimum height=0.75cm] (pbs1) at (2.2,0) {PBS};
    \node[draw,rectangle,minimum width=1.0cm,minimum height=0.75cm] (pbs2) at (7.2,0) {PBS};
    \node[above=3pt] at (2.2,0.45) {split};
    \node[above=3pt] at (7.2,0.45) {recombine};

    \draw (in) -- (pbs1.west);
    \draw (pbs2.east) -- ++(1.2,0);
    \node[above] at (8.8,0.15) {Photon out};

    \draw[double, double distance=2.5pt] ($(pbs1.east)+(0,0.24)$) -- ++(0.25,0)
        .. controls (3.05,0.24) and (3.15,0.6) .. (3.35,0.6)
        -- (6.05,0.6)
        .. controls (6.25,0.6) and (6.35,0.24) .. (6.45,0.24)
        -- ($(pbs2.west)+(0,0.24)$);
    \node[above] at (4.7,0.85) {H (no delay)};

    \node[draw,rectangle,minimum width=1.55cm,minimum height=0.75cm] (delay) at (4.3,-0.6) {delay $\Delta t$};
    \node[draw,rectangle,minimum width=0.8cm,minimum height=0.75cm] (pc) at (5.7,-0.6) {PC};
    \draw[double, double distance=2.5pt] ($(pbs1.east)+(0,-0.24)$) -- ++(0.25,0)
        .. controls (3.05,-0.24) and (3.15,-0.6) .. (delay.west);
    \draw[double, double distance=2.5pt] (delay.east) -- (pc.west);
    \draw[double, double distance=2.5pt] (pc.east) -- (6.2,-0.6)
        .. controls (6.25,-0.6) and (6.35,-0.24) .. (6.45,-0.24)
        -- ($(pbs2.west)+(0,-0.24)$);
    \node[below] at (4.7,-1.05) {V (delay $\Delta t$)};

\end{tikzpicture}
}%

\caption{PBS-delay-PBS diagram: the PBS splits H/V, the V arm experiences a delay $\Delta t$, and the second PBS recombines the paths to produce the time-bin conditioned output. PC denotes an adjustable phase controller used to calibrate the relative optical phase between the H and V arms.}

\label{fig:pbs-delay}

\end{figure}

This system implements a polarization controlled delay in the form:

\begin{equation}
\begin{aligned}
\ket{H}\ket{E} &\mapsto \ket{H}\ket{E}, \\
\ket{V}\ket{E} &\mapsto \ket{V}\ket{O}, \\
\ket{H}\ket{O} &\mapsto \ket{H}\ket{O}, \\
\ket{V}\ket{O} &\mapsto \ket{V}\ket{E},
\end{aligned}
\label{eq:pbs-delay-cnot}
\end{equation}

\subsection{Relative Optical Phases and Compensation}
\label{sec:relative-phases}

In each PBS-delay-PBS stage, the two polarization components propagate along different optical paths. To derive the resulting phase difference, we consider a narrowband photon with central vacuum wavelength $\lambda_0$ and use a common carrier reference with time dependence $e^{-i\omega_0 t}$. For a uniform fibre arm, propagation over a physical length $L_j$ contributes the phase
\begin{equation}
\phi_j = \frac{2\pi}{\lambda_0}n_{\mathrm{eff},j}L_j,
\qquad j\in\{H,V\},
\label{eq:arm-propagation-phase}
\end{equation}
where $n_{\mathrm{eff},j}$ is the effective phase refractive index of the mode propagating in arm $j$. The phase index determines the carrier phase, whereas the group index determines the pulse delay.

To display the phase factors explicitly, we suppress the temporal-mode labels in the following expressions; the H and V components still follow their respective delayed paths. For an input superposition with $|\alpha|^2+|\beta|^2=1$, the phase evolution is
\begin{equation}
\begin{aligned}
\ket{\psi_{\mathrm{in}}}
&=\alpha\ket{H}+\beta\ket{V}, \\
\ket{\psi_{\mathrm{out}}}
&=\alpha e^{i\phi_H}\ket{H}+\beta e^{i\phi_V}\ket{V} \\
&=e^{i\phi_H}\bigl(\alpha\ket{H}
+\beta e^{i(\phi_V-\phi_H)}\ket{V}\bigr).
\end{aligned}
\label{eq:path-phase-state}
\end{equation}
The common factor $e^{i\phi_H}$ is global and can be omitted. The remaining factor $e^{i(\phi_V-\phi_H)}$ multiplies only the V component, giving the relative path phase
\begin{equation}
\begin{aligned}
\phi_{\mathrm{path}}
&= \phi_V-\phi_H \\
&= \frac{2\pi}{\lambda_0}
\bigl(n_{\mathrm{eff},V}L_V-n_{\mathrm{eff},H}L_H\bigr)
\pmod{2\pi}.
\end{aligned}
\label{eq:relative-path-phase}
\end{equation}
For arms with the same effective phase index $n_{\mathrm{eff}}$, this reduces to
\begin{equation}
\phi_{\mathrm{path}}
=\frac{2\pi n_{\mathrm{eff}}\Delta L}{\lambda_0}
\pmod{2\pi},
\qquad \Delta L=L_V-L_H.
\end{equation}
This phase can occur even in a lossless, stable device. An adjustable phase element in one of the separated arms can compensate the added relative phase before recombination. For example, a phase $\phi_{\mathrm{ctrl}}$ applied to the V arm changes the relative factor in Eq.~\ref{eq:path-phase-state} to $e^{i(\phi_{\mathrm{path}}+\phi_{\mathrm{ctrl}})}$. Choosing
\begin{equation}
\phi_{\mathrm{ctrl}}=-\phi_{\mathrm{path}}\pmod{2\pi}
\label{eq:path-phase-compensation}
\end{equation}
removes the propagation-induced relative phase while preserving the relative phase of the input superposition. The phase adjustment must retain the required pulse delay. It does not itself bring the photon into another time bin.

The phases must be calibrated for the complete circuit, including the final stage, whose H arm has the additional delay $2\Delta t$. An adjustable phase element must be used for both PBS-delay-PBS stages including the final compensation stage.

Reflection and transmission in the splitting and recombination optics can add further phase offsets. These are included in the total relative arm phase when setting the adjustable phase control. The EOM may separately add a relative phase between its even-bin and odd-bin operations, even when its polarization flip is perfect. Such a phase must be characterized and, if necessary, compensated by a synchronized time-bin phase correction acting equally on both polarizations.

\subsection{Final Polarization-Dependent Compensation}

A final delay stage brings the H and V amplitudes into the same physical time bin. As shown in Fig.~\ref{fig:compensation-delay}, this stage uses the same MZI arrangement as the polarization-dependent delay, but delays H by $2\Delta t$ while leaving V unchanged. This two-bin delay preserves parity and restores the time-bin overlap within each occupied output bin, allowing the polarization amplitudes to interfere in a downstream polarization measurement.

\begin{figure}[H]

\centering

\resizebox{0.9\columnwidth}{!}{%
\begin{tikzpicture}[>=Stealth, every node/.style={font=\small}, line width=0.9pt]

    \node (in) at (0,0) {Photon in};
    \node[draw,rectangle,minimum width=1.0cm,minimum height=0.75cm] (pbs1) at (2.2,0) {PBS};
    \node[draw,rectangle,minimum width=1.0cm,minimum height=0.75cm] (pbs2) at (7.2,0) {PBS};
    \node[above=3pt] at (2.2,0.45) {split};
    \node[above=3pt] at (7.2,0.45) {recombine};

    \draw (in) -- (pbs1.west);
    \draw (pbs2.east) -- ++(1.2,0);
    \node[above] at (8.8,0.15) {Photon out};

    \draw[double, double distance=2.5pt] ($(pbs1.east)+(0,0.24)$) -- ++(0.25,0)
        .. controls (3.05,0.24) and (3.15,0.6) .. (3.35,0.6)
        -- (6.05,0.6)
        .. controls (6.25,0.6) and (6.35,0.24) .. (6.45,0.24)
        -- ($(pbs2.west)+(0,0.24)$);
    \node[above] at (4.7,0.85) {V (no delay)};

    \node[draw,rectangle,minimum width=1.55cm,minimum height=0.75cm] (delay) at (4.3,-0.6) {delay $2\Delta t$};
    \node[draw,rectangle,minimum width=0.8cm,minimum height=0.75cm] (pc) at (5.7,-0.6) {PC};
    \draw[double, double distance=2.5pt] ($(pbs1.east)+(0,-0.24)$) -- ++(0.25,0)
        .. controls (3.05,-0.24) and (3.15,-0.6) .. (delay.west);
    \draw[double, double distance=2.5pt] (delay.east) -- (pc.west);
    \draw[double, double distance=2.5pt] (pc.east) -- (6.2,-0.6)
        .. controls (6.25,-0.6) and (6.35,-0.24) .. (6.45,-0.24)
        -- ($(pbs2.west)+(0,-0.24)$);
    \node[below] at (4.7,-1.05) {H (delay $2\Delta t$)};

\end{tikzpicture}
}%

\caption{Final compensation stage: the H arm experiences a delay $2\Delta t$, while the V arm has no additional delay. PC denotes an adjustable phase controller used to calibrate the relative optical phase between the H and V arms. The second PBS recombines the paths.}

\label{fig:compensation-delay}

\end{figure}

\subsection{Full SWAP Architecture}

The three-gate sequence followed by time-delay compensation is implemented as $\mathrm{CNOT}_{P \rightarrow T}$ $\rightarrow$ $\mathrm{CNOT}_{T \rightarrow P}$ $\rightarrow$ $\mathrm{CNOT}_{P \rightarrow T}$ $\rightarrow$ H-polarization delay of $2\Delta t$, as shown in Fig.~\ref{fig:full-swap-architecture}, with V unchanged in the final delay stage.

Figure~\ref{fig:full-swap-architecture} shows the full SWAP architecture as four sequential stages: (i) a PBS-delay-PBS stage that maps polarization onto a time-bin shift, (ii) an EOM that applies the time-controlled polarization flip, (iii) a second PBS-delay-PBS stage that completes the polarization-controlled delay, and (iv) a final PBS-delay-PBS stage that delays H by $2\Delta t$ while leaving V unchanged. The photon output from each stage is routed into the next stage, so that the final output state has the polarization and time-bin qubits swapped.

\begin{figure}[H]

\centering

\resizebox{\columnwidth}{!}{%

\begin{tikzpicture}[>=Stealth, every node/.style={font=\small}, line width=0.9pt]

    \node at (4.9,1.7) {i) PBS-delay-PBS (1)};

    \node at (10.4,1.7) {ii) EOM};

    \node at (16.9,1.7) {iii) PBS-delay-PBS (2)};

    \node (in0) at (0.2,0) {Photon in};

    \node[draw,rectangle,minimum width=1.0cm,minimum height=0.75cm] (pbs1a) at (2.4,0) {PBS};
    \node[draw,rectangle,minimum width=1.0cm,minimum height=0.75cm] (pbs2a) at (7.4,0) {PBS};
    \node[draw,rectangle,minimum width=1.55cm,minimum height=0.75cm] (delaya) at (4.3,-0.6) {delay $\Delta t$};
    \node[draw,rectangle,minimum width=0.8cm,minimum height=0.75cm] (pca) at (5.7,-0.6) {PC};

    \draw[-{Stealth},shorten >=2pt] (in0) -- (pbs1a.west);

    \draw[double, double distance=2.5pt] ($(pbs1a.east)+(0,0.24)$) -- ++(0.25,0)
        .. controls (3.05,0.24) and (3.15,0.6) .. (3.35,0.6)
        -- (6.05,0.6)
        .. controls (6.25,0.6) and (6.35,0.24) .. (6.45,0.24)
        -- ($(pbs2a.west)+(0,0.24)$);
    \node[above] at (4.9,0.85) {H (no delay)};

    \draw[double, double distance=2.5pt] ($(pbs1a.east)+(0,-0.24)$) -- ++(0.25,0)
        .. controls (3.05,-0.24) and (3.15,-0.6) .. (delaya.west);
    \draw[double, double distance=2.5pt] (delaya.east) -- (pca.west);
    \draw[double, double distance=2.5pt] (pca.east) -- (6.2,-0.6)
        .. controls (6.25,-0.6) and (6.35,-0.24) .. (6.45,-0.24)
        -- ($(pbs2a.west)+(0,-0.24)$);
    \node[below] at (4.9,-1.05) {V (delay $\Delta t$)};

    \node[draw,rectangle,minimum width=1.4cm,minimum height=0.8cm] (eom) at (10.4,0) {EOM};

    \draw[->,shorten >=2pt] (pbs2a.east) -- ++(1.3,0) -- (eom.west);

    \node[draw,rectangle,minimum width=1.0cm,minimum height=0.75cm] (pbs1b) at (14.4,0) {PBS};
    \node[draw,rectangle,minimum width=1.0cm,minimum height=0.75cm] (pbs2b) at (19.4,0) {PBS};
    \node[draw,rectangle,minimum width=1.55cm,minimum height=0.75cm] (delayb) at (16.3,-0.6) {delay $\Delta t$};
    \node[draw,rectangle,minimum width=0.8cm,minimum height=0.75cm] (pcb) at (17.7,-0.6) {PC};

    \draw[-{Stealth},shorten >=2pt] (eom.east) -- (pbs1b.west);

    \draw[double, double distance=2.5pt] ($(pbs1b.east)+(0,0.24)$) -- ++(0.25,0)
        .. controls (15.05,0.24) and (15.15,0.6) .. (15.35,0.6)
        -- (18.05,0.6)
        .. controls (18.25,0.6) and (18.35,0.24) .. (18.45,0.24)
        -- ($(pbs2b.west)+(0,0.24)$);
    \node[above] at (16.9,0.85) {H (no delay)};

    \draw[double, double distance=2.5pt] ($(pbs1b.east)+(0,-0.24)$) -- ++(0.25,0)
        .. controls (15.05,-0.24) and (15.15,-0.6) .. (delayb.west);
    \draw[double, double distance=2.5pt] (delayb.east) -- (pcb.west);
    \draw[double, double distance=2.5pt] (pcb.east) -- (18.2,-0.6)
        .. controls (18.25,-0.6) and (18.35,-0.24) .. (18.45,-0.24)
        -- ($(pbs2b.west)+(0,-0.24)$);
    \node[below] at (16.9,-1.05) {V (delay $\Delta t$)};

    \begin{scope}[xshift=22cm]
        \node at (2.5,1.7) {iv) PBS-delay-PBS};
        \node[draw,rectangle,minimum width=1.0cm,minimum height=0.75cm] (comp1) at (0,0) {PBS};
        \node[draw,rectangle,minimum width=1.0cm,minimum height=0.75cm] (comp2) at (5,0) {PBS};
        \node[draw,rectangle,minimum width=1.55cm,minimum height=0.75cm] (compdelay) at (2.0,-0.6) {delay $2\Delta t$};
        \node[draw,rectangle,minimum width=0.8cm,minimum height=0.75cm] (comppc) at (3.4,-0.6) {PC};

        \draw[double, double distance=2.5pt] ($(comp1.east)+(0,0.24)$) -- (1,0.6) -- (4,0.6) -- ($(comp2.west)+(0,0.24)$);
        \node[above] at (2.5,0.85) {V (no delay)};
        \draw[double, double distance=2.5pt] ($(comp1.east)+(0,-0.24)$) -- (1,-0.6) -- (compdelay.west);
        \draw[double, double distance=2.5pt] (compdelay.east) -- (comppc.west);
    \draw[double, double distance=2.5pt] (comppc.east) -- (4,-0.6) -- ($(comp2.west)+(0,-0.24)$);
        \node[below] at (2.5,-1.05) {H (delay $2\Delta t$)};
        \draw[->,shorten >=2pt] (comp2.east) -- (7.5,0) node[midway,above] {Photon out};
    \end{scope}
    \draw[->,shorten >=2pt] (pbs2b.east) -- (comp1.west);

\end{tikzpicture}%

}

\caption{Full SWAP architecture. PC denotes an adjustable phase controller on each delayed branch: the V arms in stages (i) and (iii), and the H arm in stage (iv). These controllers calibrate the relative optical phases between the arms.}

\label{fig:full-swap-architecture}

\end{figure}

Here, we will show the state of the system's qubits at each stage of the SWAP operation. This stage-by-stage proof is meant to show the evolution of the qubit states through the optical components, and to verify that the final state corresponds to a SWAP of the polarization and time-bin qubits.

\[
\begin{aligned}
\text{We will represent:} \\
\ket{0}_P \equiv \ket{H}_P \\
\ket{1}_P \equiv \ket{V}_P \\
\ket{0}_T \equiv \ket{E}_T \\
\ket{1}_T \equiv \ket{O}_T
\end{aligned}
\]

\[
\ket{\psi_0} = (\alpha\ket{H}_P + \beta\ket{V}_P) \otimes (\gamma\ket{E}_T + \delta\ket{O}_T)
\]

\noindent\small

\[
\text{(Initial State)} \quad \ket{\psi_0} = \alpha\gamma\ket{HE} + \alpha\delta\ket{HO} + \beta\gamma\ket{VE} + \beta\delta\ket{VO}
\]

\begin{figure}[H]

\centering

\resizebox{0.72\columnwidth}{!}{%
\begin{tikzpicture}[x=1.25cm,y=0.85cm,>=Latex,font=\scriptsize]

    \draw[thick] (0,0) -- (5,0);
    \foreach \x in {0,1,2,3,4,5} {
        \draw[thick] (\x,-0.28) -- (\x,0.28);
    }

    \node[below] at (0.5,-0.28) {$E$};
    \node[below] at (1.5,-0.28) {$O$};
    \node[below] at (2.5,-0.28) {$E$};
    \node[below] at (3.5,-0.28) {$O$};
    \node[below] at (4.5,-0.28) {$E$};

    \node[align=center] at (0.5,0.95) {$\alpha\gamma\ket{HE}$\\$\beta\gamma\ket{VE}$};
    \node[align=center] at (1.5,0.95) {$\alpha\delta\ket{HO}$\\$\beta\delta\ket{VO}$};

\end{tikzpicture}%
}

\caption{Initial polarization and time-bin state.}
\label{fig:state-initial}

\end{figure}

After the first PBS-delay-PBS stage ($\mathrm{CNOT}_{P \rightarrow T}$), Vertical polarization causes a Time qubit flip, while Horizontal polarization does nothing so we have:

\[
\text{i)} \quad \ket{\psi_1} = \alpha\gamma\ket{HE} + \alpha\delta\ket{HO} + \beta\gamma\ket{VO} + \beta\delta\ket{VE}
\]

\begin{figure}[H]

\centering

\resizebox{0.72\columnwidth}{!}{%
\begin{tikzpicture}[x=1.25cm,y=0.85cm,>=Latex,font=\scriptsize]

    \draw[thick] (0,0) -- (5,0);
    \foreach \x in {0,1,2,3,4,5} {
        \draw[thick] (\x,-0.28) -- (\x,0.28);
    }

    \node[below] at (0.5,-0.28) {$E$};
    \node[below] at (1.5,-0.28) {$O$};
    \node[below] at (2.5,-0.28) {$E$};
    \node[below] at (3.5,-0.28) {$O$};
    \node[below] at (4.5,-0.28) {$E$};

    \node[align=center] at (0.5,0.95) {$\alpha\gamma\ket{HE}$};
    \node[align=center] at (1.5,0.95) {$\alpha\delta\ket{HO}$\\$\beta\gamma\ket{VO}$};
    \node[align=center] at (2.5,0.95) {$\beta\delta\ket{VE}$};

\end{tikzpicture}%
}

\caption{Time-bin state after the first PBS-delay-PBS stage (i).}
\label{fig:state-first-delay}

\end{figure}

After the EOM stage ($\mathrm{CNOT}_{T \rightarrow P}$), the time-bin $\ket{O}_T$ is aligned with the EOM voltage so it causes a polarization flip, while the time-bin $\ket{E}_T$ is unaffected so we have:

\[
\text{ii)} \quad \ket{\psi_2} = \alpha\gamma\ket{HE} + \alpha\delta\ket{VO} + \beta\gamma\ket{HO} + \beta\delta\ket{VE}
\]

\begin{figure}[H]

\centering

\resizebox{0.72\columnwidth}{!}{%
\begin{tikzpicture}[x=1.25cm,y=0.85cm,>=Latex,font=\scriptsize]

    \draw[thick] (0,0) -- (5,0);
    \foreach \x in {0,1,2,3,4,5} {
        \draw[thick] (\x,-0.28) -- (\x,0.28);
    }

    \node[below] at (0.5,-0.28) {$E$};
    \node[below] at (1.5,-0.28) {$O$};
    \node[below] at (2.5,-0.28) {$E$};
    \node[below] at (3.5,-0.28) {$O$};
    \node[below] at (4.5,-0.28) {$E$};

    \node[align=center] at (0.5,0.95) {$\alpha\gamma\ket{HE}$};
    \node[align=center] at (1.5,0.95) {$\alpha\delta\ket{VO}$\\$\beta\gamma\ket{HO}$};
    \node[align=center] at (2.5,0.95) {$\beta\delta\ket{VE}$};

\end{tikzpicture}%
}

\caption{Time-bin state after the EOM stage (ii).}
\label{fig:state-eom}

\end{figure}

After the second PBS-delay-PBS stage ($\mathrm{CNOT}_{P \rightarrow T}$), again Vertical polarization causes a Time qubit flip, while Horizontal polarization does nothing so we have:

\[
\text{iii)} \quad \ket{\psi_3} = \alpha\gamma\ket{HE} + \alpha\delta\ket{VE} + \beta\gamma\ket{HO} + \beta\delta\ket{VO}
\]

\begin{figure}[H]

\centering

\resizebox{0.72\columnwidth}{!}{%
\begin{tikzpicture}[x=1.25cm,y=0.85cm,>=Latex,font=\scriptsize]

    \draw[thick] (0,0) -- (5,0);
    \foreach \x in {0,1,2,3,4,5} {
        \draw[thick] (\x,-0.28) -- (\x,0.28);
    }

    \node[below] at (0.5,-0.28) {$E$};
    \node[below] at (1.5,-0.28) {$O$};
    \node[below] at (2.5,-0.28) {$E$};
    \node[below] at (3.5,-0.28) {$O$};
    \node[below] at (4.5,-0.28) {$E$};

    \node[align=center] at (0.5,0.95) {$\alpha\gamma\ket{HE}$};
    \node[align=center] at (1.5,0.95) {$\beta\gamma\ket{HO}$};
    \node[align=center] at (2.5,0.95) {$\alpha\delta\ket{VE}$};
    \node[align=center] at (3.5,0.95) {$\beta\delta\ket{VO}$};

\end{tikzpicture}%
}

\caption{Time-bin state after the second PBS-delay-PBS stage (iii).}
\label{fig:state-second-delay}

\end{figure}

After stage (iii), the parity labels have the desired logical arrangement, but the relevant H and V amplitudes still occupy different physical bins, as shown in Fig.~\ref{fig:state-second-delay}. In general, the physical state cannot yet be factorized into the desired polarization and time-bin output states.

After the final PBS-delay-PBS stage (iv), the H-polarized components are delayed by $2\Delta t$, while the V-polarized components remain unchanged. The relevant H and V amplitudes then overlap within each output time bin, as shown in Fig.~\ref{fig:state-compensated}.

\begin{figure}[H]

\centering

\resizebox{0.72\columnwidth}{!}{%
\begin{tikzpicture}[x=1.25cm,y=0.85cm,>=Latex,font=\scriptsize]

    \draw[thick] (0,0) -- (5,0);
    \foreach \x in {0,1,2,3,4,5} {
        \draw[thick] (\x,-0.28) -- (\x,0.28);
    }
    \node[below] at (0.5,-0.28) {$E$};
    \node[below] at (1.5,-0.28) {$O$};
    \node[below] at (2.5,-0.28) {$E$};
    \node[below] at (3.5,-0.28) {$O$};
    \node[below] at (4.5,-0.28) {$E$};

    \node[align=center] at (2.5,0.95) {$\alpha\gamma\ket{HE}$\\$\alpha\delta\ket{VE}$};
    \node[align=center] at (3.5,0.95) {$\beta\gamma\ket{HO}$\\$\beta\delta\ket{VO}$};

\end{tikzpicture}%
}

\caption{Time-bin state after the final H-polarization delay of $2\Delta t$ (iv). The H and V amplitudes overlap within each occupied time-bin.}
\label{fig:state-compensated}

\end{figure}

With the relative optical phases calibrated, the compensated state can now be factorized as
\[
\begin{aligned}
\ket{\psi_4}
&=\alpha\gamma\ket{HE}+\alpha\delta\ket{VE}
 +\beta\gamma\ket{HO}+\beta\delta\ket{VO} \\
&=(\gamma\ket{H}_P+\delta\ket{V}_P)
\otimes(\alpha\ket{E}_T+\beta\ket{O}_T).
\end{aligned}
\]
Here, $\ket{E}_T$ and $\ket{O}_T$ denote the common output even/odd pair shown in Fig.~\ref{fig:state-compensated}, located $2\Delta t$ later than the corresponding input pair. Thus, the complete compensated circuit exchanges the polarization and time-bin qubit states while translating the pair forward in time.

\medskip

\section{Experimental Constraints}
\label{sec:constraints}

\subsection{Timing Requirements}

The time-bin separation $\Delta t$ must satisfy: $\Delta t \gg \tau_{switch}$ and $\Delta t > \tau_{open}$. Equivalently, a timing budget can be written as $\Delta t \gtrsim \tau_{open} + \tau_{switch}$, so each interval includes both transition time and a stable sampling region. where $\tau_{switch}$ is the EOM switching time and $\tau_{open}$ is the usable flat sampling window. The time-bin separation $\Delta t$ must be long enough to allow the EOM to switch between voltage levels and to provide a stable sampling window for the photon detection.

To avoid ambiguity, we use the same $\Delta t$ for both (i) the photon time-bin spacing and (ii) the synchronized EOM timing interval. The quantity $\tau_{switch}$ denotes the EOM transition (rise/fall) time. We denote the usable flat sampling window by $\tau_{open}$.

Figure~\ref{fig:eom-eye} illustrates a finite-rise/fall EOM drive as an eye diagram. The interval from $t_n$ to $t_n+\Delta t$ is taken between adjacent crossing points of the eye, and the valid sampling window is the central flat portion of the eye opening.

\begin{figure}[H]

\centering

\resizebox{0.9\columnwidth}{!}{%
\begin{tikzpicture}[x=1.25cm,y=1.05cm,>=Latex]

        \draw[->] (-0.1,0) -- (7.8,0) node[right] {$t$};
        \draw[->] (0,-0.15) -- (0,2.0) node[above] {EOM voltage};

        \draw[densely dashed,gray] (2.0,0) -- (2.0,1.95);
        \draw[densely dashed,gray] (4.0,0) -- (4.0,1.95);
        \node[below] at (2.0,0) {$t_n$};
        \node[below] at (4.0,0) {$t_n+\Delta t$};

        \draw[decorate,decoration={brace,amplitude=4pt}] (2.0,-0.28) -- (4.0,-0.28)
            node[midway,yshift=-10pt] {$\Delta t$ (eye width)};

        \draw[densely dotted,gray] (0.2,1.62) -- (7.6,1.62);
        \draw[densely dotted,gray] (0.2,0.38) -- (7.6,0.38);
        \node[left] at (0.2,1.62) {high};
        \node[left] at (0.2,0.38) {low};

        \draw[thick,blue]
            (0.25,0.38)
            -- (1.55,0.38)
            .. controls (1.74,0.38) and (1.90,1.00) .. (2.00,1.00)
            .. controls (2.10,1.00) and (2.26,1.62) .. (2.45,1.62)
            -- (3.55,1.62)
            .. controls (3.74,1.62) and (3.90,1.00) .. (4.00,1.00)
            .. controls (4.10,1.00) and (4.26,0.38) .. (4.45,0.38)
            -- (5.55,0.38)
            .. controls (5.74,0.38) and (5.90,1.00) .. (6.00,1.00)
            .. controls (6.10,1.00) and (6.26,1.62) .. (6.45,1.62)
            -- (7.45,1.62);

        \draw[thick,blue]
            (0.25,1.62)
            -- (1.55,1.62)
            .. controls (1.74,1.62) and (1.90,1.00) .. (2.00,1.00)
            .. controls (2.10,1.00) and (2.26,0.38) .. (2.45,0.38)
            -- (3.55,0.38)
            .. controls (3.74,0.38) and (3.90,1.00) .. (4.00,1.00)
            .. controls (4.10,1.00) and (4.26,1.62) .. (4.45,1.62)
            -- (5.55,1.62)
            .. controls (5.74,1.62) and (5.90,1.00) .. (6.00,1.00)
            .. controls (6.10,1.00) and (6.26,0.38) .. (6.45,0.38)
            -- (7.45,0.38);

        \draw[blue] (1.90,0.90) -- (2.10,1.10);
        \draw[blue] (1.90,1.10) -- (2.10,0.90);
        \draw[blue] (3.90,0.90) -- (4.10,1.10);
        \draw[blue] (3.90,1.10) -- (4.10,0.90);

        \fill[green!18] (2.45,0.62) rectangle (3.55,1.38);
        \draw[densely dashed,green!60!black] (2.45,0.55) -- (2.45,1.70);
        \draw[densely dashed,green!60!black] (3.55,0.55) -- (3.55,1.70);
        \node[align=center,font=\scriptsize] at (3.00,1.00) {sampling\\window};

        \fill[red] (3.00,1.62) circle (1.8pt);
        \fill[red] (3.00,0.38) circle (1.8pt);
        \draw[->,red] (3.00,1.90) -- (3.00,1.66);
        \draw[->,red] (3.00,0.10) -- (3.00,0.34);
\end{tikzpicture}%
}

    \caption{EOM timing eye with finite rise/fall edges. The eye width is defined by adjacent crossing points $t_n$ and $t_n+\Delta t$, while the valid sampling window is the central flat part of the eye opening.}
\label{fig:eom-eye}

\end{figure}

\subsection{EOM Rotation Accuracy Limits}

For $\mathrm{CNOT}_{T \rightarrow P}$, the odd time-bin polarization flip should approximate a half-wave action in the $H/V$ basis. We model the EOM as a linear retarder with retardance $\theta$, whose principal axes are oriented at $45^\circ$ relative to the $H/V$ basis. Using the Jones-calculus description of such a retarder \cite{ref_hecht_optics}, we obtain the following associated transition probabilities:

\begin{equation}
\begin{aligned}
P_{H\rightarrow H}
&=\left|A_{H\rightarrow H}\right|^2
=\cos^2\!\left(\frac{\theta}{2}\right), \\
P_{H\rightarrow V}
&=\left|A_{H\rightarrow V}\right|^2
=\sin^2\!\left(\frac{\theta}{2}\right).
\end{aligned}
\end{equation}

Defining the polarization-flip probability as $P_{\mathrm{flip}}=P_{H\rightarrow V}$, we obtain $P_{\mathrm{flip}}=\sin^2(\theta/2)$. The same flip probability applies to a V-polarized input. Using $\theta=\pi V/V_{\pi}$ gives
\begin{equation}
P_{\mathrm{flip}}=\sin^2\!\left(\frac{\pi V}{2V_{\pi}}\right).
\end{equation}

For the intended half-wave operation, write $V=V_{\pi}(1+\epsilon_V)$, where $\epsilon_V$ is the fractional drive-voltage error. The successful flip probability is then
\begin{equation}
\begin{aligned}
P_{\mathrm{success}}=P_{\mathrm{flip}}
&=\sin^2\!\left(\frac{\pi}{2}+\frac{\pi\epsilon_V}{2}\right) \\
&=\cos^2\!\left(\frac{\pi\epsilon_V}{2}\right).
\end{aligned}
\label{eq:eom-flip-success}
\end{equation}
In this lossless model, the probability of failing to flip an H/V basis input is therefore
\begin{equation}
P_{\mathrm{error}}=1-P_{\mathrm{success}}
=\sin^2\!\left(\frac{\pi\epsilon_V}{2}\right).
\label{eq:eom-flip-error}
\end{equation}
For $|\epsilon_V|\ll1$, this becomes
\begin{equation}
P_{\mathrm{error}}\approx\frac{\pi^2\epsilon_V^2}{4}.
\label{eq:eom-flip-error-small}
\end{equation}
Thus, small voltage deviations produce a flip-failure probability that is quadratic in the fractional voltage error. At $\epsilon_V=0$, the success probability is unity and the error probability vanishes.

These probabilities characterize polarization-flip outcomes for H/V basis inputs in the odd bins; they do not establish the fidelity of the full quantum gate. Even a perfect flip can be accompanied by an incorrect phase relative to the even-bin operation, changing the output for a superposition of even and odd bins. Coherent gate performance therefore also requires characterization of relative phases and temporal-mode overlap, as discussed in Sec.~\ref{sec:relative-phases}.

\subsection{Choosing $\Delta t$ From EOM and Detector Resolution}

In practice, $\Delta t$ should be selected from the slower of two timing limits: (i) EOM transitions and settling and (ii) detection timing resolution. A useful engineering rule is

\begin{equation}
\Delta t \gtrsim \tau_{switch} + \tau_{open} + \tau_{det} + \sigma_{jitter},
\end{equation}

\noindent where $\tau_{det}$ is the effective detector/electronics timing aperture and $\sigma_{jitter}$ is the combined RMS timing jitter of the source trigger, EOM drive, and detector chain. Here, $\sigma_{jitter}$ is included as a representative timing allowance rather than as a specified statistical confidence margin.

This makes the technology question explicit: once representative values of $\tau_{switch}$, $\tau_{open}$, $\tau_{det}$, and $\sigma_{jitter}$ are chosen for the available hardware, the minimum viable $\Delta t$ follows directly. A more conservative design may include an additional safety factor for the measured timing jitter.

Prior demonstrations provide realistic values for these timing parameters. For an electro-optic photonic switch, Švarc \emph{et al.} reported a $10\,\mathrm{GHz}$ electrical bandwidth and subnanosecond switching, with the switching rise time estimated to be below $100\,\mathrm{ps}$ \cite{ref_svarc_eom}.

Time-bin experiments can also span subnanosecond to nanosecond-scale intervals. For example, Ikuta \emph{et al.} analyzed time-bin--polarization conversion using a $200\,\mathrm{ps}$ time bin \cite{ref_ikuta_interface}, while Sun \emph{et al.} used approximately $1\,\mathrm{ns}$ temporal separations in a fiber-based time-bin experiment \cite{ref_sun_1ns}. For the present architecture, a practical first design point is therefore $\Delta t=1\,\mathrm{ns}$. Taking $\tau_{switch}=100\,\mathrm{ps}$, $\tau_{open}=300\,\mathrm{ps}$, $\tau_{det}=100\,\mathrm{ps}$, and $\sigma_{jitter}=50\,\mathrm{ps}$ gives a total nominal budget of $550\,\mathrm{ps}$ and leaves approximately $450\,\mathrm{ps}$ of timing margin. If four timing slots are allocated to the complete three-CNOT sequence, the minimum overall gate time is approximately $4\times550\,\mathrm{ps}=2.2\,\mathrm{ns}$. Allowing additional inter-stage overhead and rounding to a practical value gives an overall gate time of approximately $2.5\,\mathrm{ns}$, corresponding to an ideal average gate rate of $R_{\mathrm{gate}}\approx1/(2.5\,\mathrm{ns})\approx400\,\mathrm{MHz}$ before accounting for optical loss, detector dead time, and other experimental inefficiencies.

\section{Discussion and Implications}

The proposed circuit could provide a physical realization of the polarization--time-bin SWAP required in the hyperentangled Bell-state analyzer of Zhang \emph{et al.}~\cite{ref_zhang_hyperentangled_analysis}. In that protocol, the SWAP transfers time-bin information into polarization for subsequent analysis using polarization parity-check quantum nondemolition measurements. 

The same CNOT operations can be useful beyond routing. When a photon carries logical qubits in more than one degree of freedom, polarization--time-bin CNOTs can couple those qubits and could serve as local gate primitives for parity checks in multidegree-of-freedom entanglement-purification protocols. Such protocols would also require suitable shared entangled states, measurements, and classical communication. Their design and analysis are beyond the scope of the present work.

Several limitations of the present analysis should be noted. The scheme is presented at the unitary level and does not yet include a quantitative hardware-level analysis for (i) finite EOM rotation accuracy versus drive-voltage error and bandwidth limits, and (ii) detector/electronics timing resolution and jitter limitations versus the chosen $\Delta t$. A technology-specific budget should be constructed to verify that the selected $\Delta t$ leaves a safe sampling margin.

\section{Conclusion}

We have presented a new polarization--time-bin SWAP circuit that combines bidirectional parity changes produced by fixed polarization-dependent delays with synchronized electro-optic modulation. A final temporal-compensation stage aligns the relevant polarization amplitudes in common output bins, enabling coherent exchange of the polarization and time-bin qubit states when the required phase conditions are satisfied. Even and odd labels describe the repeating early/late time-bin grid used to track the temporal routing. The proposed circuit and accompanying analysis provide design criteria for implementing a coherent SWAP between polarization and time-bin qubits, including temporal recombination, optical phase compensation, and voltage and timing tolerances. The circuit provides a routing primitive for multi-degree-of-freedom photonic systems and a candidate physical realization of the polarization--time-bin SWAP used in hyperentangled-state analysis~\cite{ref_zhang_hyperentangled_analysis}.

\end{document}